\documentclass[superscriptaddress,showpacs,amsmath,amssymb,aps,pra,twocolumn]{revtex4-1}
\usepackage{graphicx}
\usepackage{dcolumn}
\usepackage{bm}
\usepackage{hyperref}
\usepackage{color}
\usepackage{textcomp}
\newcommand{\tabincell}[2]{\begin{tabular}{@{}#1@{}}#2\end{tabular}}

\begin{document}
\title{Feshbach spectroscopy of an ultracold $^{41}$K-$^6$Li mixture and $^{41}$K atoms}

\author{Xiang-Pei Liu}
\affiliation{Shanghai Branch, National Laboratory for Physical Sciences at Microscale and Department of Modern Physics, University of Science and Technology of China, Shanghai, 201315, China}
\affiliation{CAS Center for Excellence and Synergetic Innovation Center in Quantum Information and Quantum Physics, University of Science and Technology of China, Hefei, Anhui 230026, China}
\affiliation{CAS-Alibaba Quantum Computing Laboratory, Shanghai, 201315, China}

\author{Xing-Can Yao}
\affiliation{Shanghai Branch, National Laboratory for Physical Sciences at Microscale and Department of Modern Physics, University of Science and Technology of China, Shanghai, 201315, China}
\affiliation{CAS Center for Excellence and Synergetic Innovation Center in Quantum Information and Quantum Physics, University of Science and Technology of China, Hefei, Anhui 230026, China}
\affiliation{CAS-Alibaba Quantum Computing Laboratory, Shanghai, 201315, China}

\author{Ran Qi}
\affiliation{Department of Physics, Renmin University of China, Beijing, 100872, China}

\author{Xiao-Qiong Wang}
\affiliation{Shanghai Branch, National Laboratory for Physical Sciences at Microscale and Department of Modern Physics, University of Science and Technology of China, Shanghai, 201315, China}
\affiliation{CAS Center for Excellence and Synergetic Innovation Center in Quantum Information and Quantum Physics, University of Science and Technology of China, Hefei, Anhui 230026, China}
\affiliation{CAS-Alibaba Quantum Computing Laboratory, Shanghai, 201315, China}

\author{Yu-Xuan Wang}
\affiliation{Shanghai Branch, National Laboratory for Physical Sciences at Microscale and Department of Modern Physics, University of Science and Technology of China, Shanghai, 201315, China}
\affiliation{CAS Center for Excellence and Synergetic Innovation Center in Quantum Information and Quantum Physics, University of Science and Technology of China, Hefei, Anhui 230026, China}
\affiliation{CAS-Alibaba Quantum Computing Laboratory, Shanghai, 201315, China}

\author{Yu-Ao Chen}
\affiliation{Shanghai Branch, National Laboratory for Physical Sciences at Microscale and Department of Modern Physics, University of Science and Technology of China, Shanghai, 201315, China}
\affiliation{CAS Center for Excellence and Synergetic Innovation Center in Quantum Information and Quantum Physics, University of Science and Technology of China, Hefei, Anhui 230026, China}
\affiliation{CAS-Alibaba Quantum Computing Laboratory, Shanghai, 201315, China}

\author{Jian-Wei Pan}
\affiliation{Shanghai Branch, National Laboratory for Physical Sciences at Microscale and Department of Modern Physics, University of Science and Technology of China, Shanghai, 201315, China}
\affiliation{CAS Center for Excellence and Synergetic Innovation Center in Quantum Information and Quantum Physics, University of Science and Technology of China, Hefei, Anhui 230026, China}
\affiliation{CAS-Alibaba Quantum Computing Laboratory, Shanghai, 201315, China}

\begin{abstract}
We have observed 69 $^{41}$K-$^6$Li interspecies Feshbach resonances including 13 elastic p-wave resonances and 6 broad d-wave resonances of $^{41}$K atoms in different spin-state combinations at fields up to 600~G. Multi-channel quantum defect theory calculation is performed to assign these resonances and the results show perfect agreement with experimental values after improving input parameters. The observed broad p- and d- wave resonances display a full resolved multiplet structure. They may serve as important simulators to nonzero partial wave dominated physics.
\end{abstract}

\maketitle

\section{Introduction}
Magnetic Feshbach resonance~\cite{Chin2010RMP} provides an essential tool in the study of ultracold atom physics. Several breakthroughs have been demonstrated by tuning the interaction strength with Feshbach resonance, such as the formation of solitons~\cite{Khaykovich2002S,Strecker2002N}, the observation of Efimov trimer states~\cite{Kraemer2006N,Zaccanti2009NP,Knoop2009NP}, and the creation of polar molecules~\cite{Ni2008S}. Of particular interests are the broad Feshbach resonances, which are useful for studying universal properties in few-body and many-body atomic systems. Broad s-wave Feshbach resonance offers a great opportunity to study both weak-coupling Bardeen-Cooper-Schrieffer (BCS) superfluid and Bose-Einstein condensation (BEC) of tightly bound fermion pairs~\cite{Regal2004PRL,Bourdel2004PRL,Chin2004S,Zwierlein2005N}. Broad high partial wave resonances hold great promise for the study of three-body recombination decay~\cite{Suno2003NJoP}, the formation of p-wave molecules~\cite{Gaebler2007PRL}, and the realization of d-wave molecular superfluid~\cite{Yao2017}.

\begin{figure}[htbp]
\centering
\includegraphics[width=0.48\textwidth]{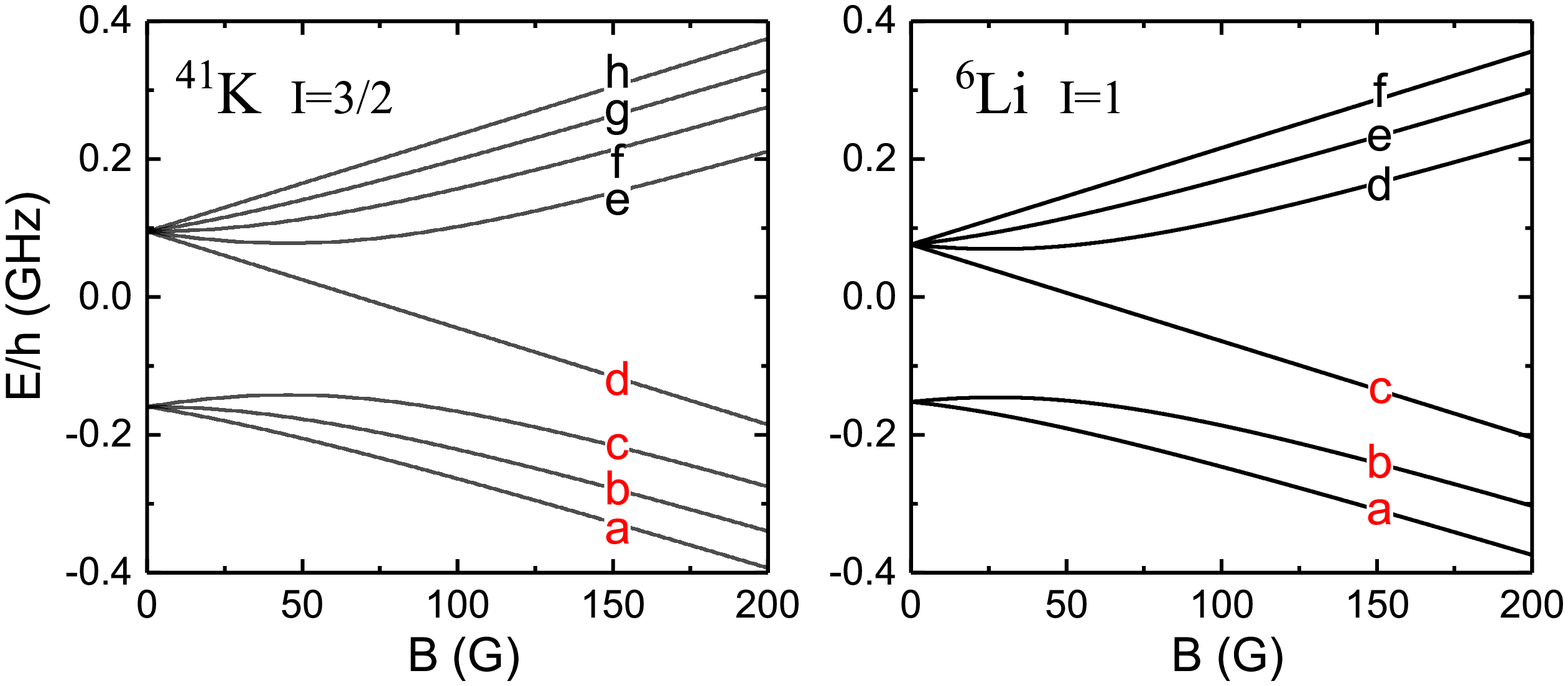}
\caption{The magnetic field dependence of the atomic ground state energies of $^{41}$K and $^6$Li in different zeeman levels. The levels of interest are marked as red. Throughout this paper, we use channel $|ab\rangle$ to indicate $^{41}$K in $|a\rangle$ and $^6$Li in $|b\rangle$, or $^{41}$K in $|ab\rangle$ mixture, and so on.}
\label{level}
\end{figure}

Up to now, Feshbach resonances have been studied in most of the laser-cooled atomic species~\cite{Chin2010RMP,Frisch2014N,Maier2015PRA}. Among them, lithium and potassium atoms have attracted intense interests both in experimental and theoretical studies. Many of s- or p- wave resonances have been reported in the isotopes of lithium and potassium~\cite{Jochim2002PRL,Khaykovich2002S,DErrico2007NJP,Loftus2002PRL,Chen2016PRA}, especially two broad s-wave resonances in $^6$Li~\cite{Jochim2003S} and $^{40}$K~\cite{Greiner2003N}, respectively. Interspecies Feshbach resonances have been studied in the Fermi-Fermi mixture of $^6$Li and $^{40}$K~\cite{Wille2008PRL,Tiecke2010PRL}. The potential functions of singlet and triplet states of lithium-potassium mixture have also been observed~\cite{Tiemann2009PRA}. For the Bose-Fermi mixture of $^{41}$K-$^6$Li, although superfluid mixture have been realized with the observation of vortex lattices~\cite{Yao2016PRL}, a full description of its Feshbach resonance is still missing. Moreover, the recent observation of a broad d-wave shape resonance in the lowest energy channel of $^{41}$K atoms~\cite{Yao2017} brings great promise for the investigation of universal physics with strong coupling in nonzero partial waves. Therefore, a more complete search on the d-wave resonances in $^{41}$K systems is urgently required.

In this work, we report on the first extensive experimental and theoretical study on the s- and p- wave Feshbach resonances in $^{41}$K-$^6$Li mixture and d-wave Feshbach resonances in single species $^{41}$K. We observe 69 interspecies Feshbach resonances in 10 spin combinations of $^{41}$K-$^6$Li mixture and 6 new broad d-wave resonances in 4 spin combinations of $^{41}$K atoms (see Fig.~\ref{level} for the interested spin channels). We perform semi-analytic multi-channel quantum defect theory (MQDT) calculation based on parameters reported in previous literatures~\cite{Tiemann2009PRA} to identify these resonances. After fine tuning of the singlet and
triplet quantum defect parameters, perfect accordance are achieved between the theoretical and experimental results. In the case of interspecies Feshbach resonances, all the observed s-wave resonances are identified as narrow resonances, while several p-wave resonances possess quite wide resonance widths, which are also experimentally ascertained by the distinct doublet structure on the loss spectroscopy. For the single species $^{41}$K system, all the d-wave broad resonances possess fully resolved triplet structure on the loss spectroscopy, which exactly demonstrates that they are from real d-wave to d-wave coupling~\cite{Yao2017,Cui2017PRL}. Our vast amounts of experimental data and improved scattering parameters are important to the future research on $^{41}$K-$^6$Li mixture and $^{41}$K system.

This paper will be organized as following: In section II, we briefly introduce the experiment method for preparing the ultracold $^{41}$K and $^6$Li atoms and detecting the scattering resonances therein; In section III, the details of MQDT calculation are given; In section IV, we present the results of s-wave and p-wave resonances in $^{41}$K-$^6$Li mixture, and d-wave resonances in single $^{41}$K system, respectively; In section V, a conclusion and outlook are drawn.

\section{Experimental procedure}
The experimental procedure to produce the ultracold $^{41}$K-$^6$Li mixture has been described in our previous works~\cite{Chen2016PRA,Yao2016PRL,Wu2017JPB}. After radio frequency (rf) evaporative cooling in the optically-plugged magnetic trap, we load about $9\times10^6$ $^6$Li and $3\times10^6$ $^{41}$K atoms into a cigar-shaped optical dipole trap and  immediately prepare them in the lowest spin states, respectively. Then, we increase the magnetic field to 435~G and implement a 8~s forced evaporative cooling on $^{41}$K atoms by lowering the laser intensity, while the $^6$Li atoms are sympathetically cooled. Next, the cold mixture is adiabatically transferred into a disk-like optical dipole trap, where two elliptical laser beams with aspect ratio of 4:1 are crossed perpendicularly (wavelength 1064~nm, maximum laser power of each beam 1.1~W, $1/e^2$ axial (radial) radius 48~$\mu$m (200~$\mu$m)). Further evaporative cooling of 8~s is performed within the disk-like trap, where ultracold $^{41}$K-$^6$Li mixture with targeted atom numbers and temperatures can be prepared by choosing appropriate final laser intensity.  Finally, absorption imaging along the gravity direction is applied to probe the two species in a single experimental run.

For the s-wave Feshbach resonance measurement, we prepare the mixture at lowest temperature with $2.2\times10^5$ $^{41}$K atoms at 60~nK or 0.65~BEC temperature ($T_c$) and $5.5\times10^5$ $^6$Li atoms at 270~nK or 0.5~Fermi temperature ($T_F$), respectively, to reduce the temperature induced broaden and resonance position shift of the loss spectrum. We mention that the $^{6}$Li and $^{41}$K atoms cannot reach fully thermal equilibrium because the single spin $^{6}$Li atoms can only thermalize through collisions with  $^{41}$K atoms and the cloud size difference between them is huge. Fortunately, we find this effect does not affect the Feshbach spectroscopy. For the p-wave resonance measurements, to allow atoms tunnelling through the centrifugal barrier, we prepare the mixture at higher temperatures with $1.4\times10^6$ $^{41}$K atoms at 350~nK and $1\times10^6$ $^6$Li atoms at 480~nK, respectively. Next, to prepare different spin state combinations, several Landau-Zener pulses are successively applied to transfer the $^6$Li $|a\rangle$ and $^{41}$K $|a\rangle$ atoms to the desired spin states with efficiencies higher than 99\%.

To facilitate the measurements, we use the same experimental sequence to produce the ultracold $^{41}$K atoms, but without loading $^{6}$Li atoms at the magneto-optical trap stage. In order to detect the triplet splitting of d-wave resonances, the cloud is prepared at 310~nK, which is slightly below the $T_c$~\cite{Yao2017} with more than $1.8\times 10^6$ $^{41}$K atoms. Then, we fine tune the rf power of Landau-Zener pulses to prepare the targeted spin combinations of $^{41}$K atoms. We mention that the background lifetime for all the spin combinations of $^{41}$K-$^6$Li mixture and pure $^{41}$K are more than 10~s, even for the inelastic scattering channels, long enough for the following measurements.

\begin{figure}[htbp]
\includegraphics[width=0.48\textwidth]{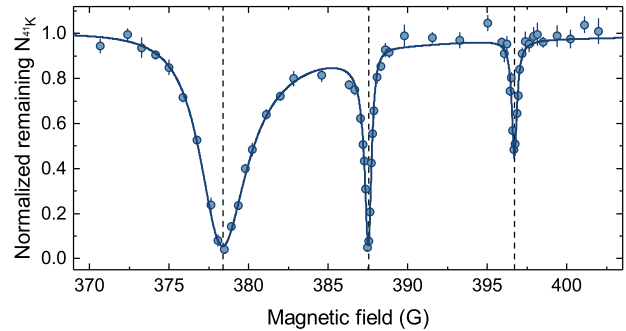}
\caption{Normalized remaining atom number of $^{41}$K as a function of magnetic field, near three s-wave resonances in $^{41}$K-$^6$Li channel $|ca\rangle$. The solid line is the multi-peak asymmetric Lorentz fitting curve. The vertical dashed lines mark the fitted resonance positions. All the data points are averaged out of five records and the error bars represent the standard deviations.}
\label{swave}
\end{figure}

After initial state preparation, inelastic loss spectroscopy is performed to detect the scattering resonances where enhanced atom losses occur due to three-body decay. For each measurement, to reduce the systematic errors, the magnetic field is ramped from 435 G to a few Gauss below or above the resonance in 50~ms, depending on the resonance position is  below or above 435 G, respectively. Then, after a 100~ms holding time at that magnetic filed for equilibrium, the magnetic field is quickly switched to a desired value and hold for some time between 100~ms to 5~s. The waiting time is carefully chosen for different scattering resonances such that obvious atom loss signal can be acquired and detectable atoms still survive in the trap to reduce the saturation effects. Finally, the magnetic field is ramped back to 435~G to simultaneously detect the remaining $^6$Li and $^{41}$K atom number using resonant absorption imaging. It's found that the loss signals of $^6$Li atoms are not as good as $^{41}$K atoms, mainly due to the large difference of cloud sizes between $^6$Li and $^{41}$K cloud. Therefore, we only adopt the results of $^{41}$K atoms to derive the parameters of scattering resonances. Moreover, additional measurements on single-species $^6$Li and $^{41}$K atoms are respectively taken to ensure the observed loss spectrums are solely contributed by the interspecies interactions.

\section{Theoretical prediction of Feshbach Resonances}
To predict the position and other scattering parameters of the Feshbach Resonances, we adopt the semi-analytical MQDT theory developed in \cite{Gao1998PRA,Gao2005PRA,Gao2008PRA}. In this approach, two quantum defect functions $K^c_S(\epsilon,\ell),~K^c_T(\epsilon,\ell)$ are required as inputs where $\epsilon,~\ell$ are scattering energy and angular momentum, respectively. When completely neglecting the dependence on $\epsilon,~\ell$ and setting $K^c_S(\epsilon,\ell)=K^c_S(0,0),~K^c_T(\epsilon,\ell)=K^c_T(0,0)$, one can already obtain most of the resonance positions with reasonable agreement comparing to the experimental results \cite{Cui2017PRL,Makrides2014PRA}. However, as pointed out in \cite{Makrides2014PRA}, including the $\ell$ dependence with a linear form in $K^c_S(\epsilon,\ell),~K^c_T(\epsilon,\ell)$ greatly improves the accuracy and extends the predicting power of this MQDT.

Considering the correction from the $\epsilon$ dependence is usually much smaller comparing to that from $\ell$ dependence, we choose the following form in the study of $^{41}$K-$^6$Li resonances:
\begin{eqnarray}
K^c_S(\epsilon,\ell)=K^c_S(0,0)+\beta_S\ell(\ell+1),\label{K_S}\\
K^c_S(\epsilon,\ell)=K^c_T(0,0)+\beta_T\ell(\ell+1).\label{K_T}
\end{eqnarray}
With such choice of parametrization for $K^c_S(\epsilon,\ell),~K^c_T(\epsilon,\ell)$, all scattering properties in all partial waves and all different hyperfine levels, can be determined by four parameters $K^c_S(0,0),~K^c_T(0,0)$ and $\beta_S,~\beta_T$. In turn, these four parameters can be fixed with four different known resonance positions in two different $\ell$ (two resonances for each $\ell$) from experiments. In practice, we choose two s-wave resonance at $B=352.46~G$~(channel $|ba\rangle$) and $B=410.54~G$~(channel $|bc\rangle$) and two p-wave resonance at $B=210.09~G$~(channel $|ca\rangle$) and $B=239.71~G$~(channel $|cb\rangle$) to fix these four parameters. As a starting point, we first relate the parameters $K^c_{S,T}(0,0)$ to the singlet and triplet scattering length $a_{S},~a_T$ through the analytic formula \cite{Gao2005PRA,Hanna2009PRA}:
\begin{eqnarray}
\frac{a_{S,T}}{\bar{a}}=\sqrt{2}\frac{K^c_{S,T}(0,0)+\tan(\pi/8)}{K^c_{S,T}(0,0)-\tan(\pi/8)}.
\end{eqnarray}
If we take $a_S=42.75a_0,~a_T=60.77a_0$ from early literature~\cite{Tiemann2009PRA}, we get $K^c_{S}(0,0)=-3.2704,~K^c_{T}(0,0)=8.5430$. Taking these values of $K^c_{S,T}(0,0)$ to calculate the s-wave scattering, the predicted resonance positions will deviate from the experimental values from several Gauss to dozens of Gauss. Fine tunings of $K^c_{S,T}(0,0)$ are then necessary to get a better fit. Finally, we get $K^c_S(0,0)=-3.1235$ and $K^c_T(0,0)=8.9785$, which results in the new values $a_S=42.24a_0$, $a_T=60.48a_0$. Under these choices, the predicted s-wave resonance positions agree perfectly with experimental results, as shown in Table~\ref{table:swave}.

For the $^{41}$K-$^6$Li p-wave resonances, the predicted resonance positions with fine tuned $K^c_{S,T}(0,0)$ still deviate from the experimental values~\cite{respos} on the level of ten Gauss, if we simply neglect the $\ell$ dependence in $K^c_S(\epsilon,\ell)$ and set $\beta_{S}=\beta_T=0$ in Eq. (\ref{K_S}) and (\ref{K_T}). Then by further adjusting the values of $\beta_{S,T}$ to $~\beta_S=0.0455$, $\beta_T=0.3407$, we can also get perfect agreement for the p-wave resonances between theory and experimental results, as shown in Table~\ref{table:pwave}. As for the $^{41}K$ resonances, since we focus on the d-wave resonances in this work, we only need two parameters $K^c_S(0,\ell=2)$ and $K^c_T(0,\ell=2)$ which is fixed by two d-wave resonance positions. By choosing the two resonances in the channel~$|aa\rangle$ (see Table~\ref{table:dwave}), we find $K^c_S(0,\ell=2)=2.0985,~K^c_T(0,\ell=2)=11.2949$. Similarly, one can also see the excellent quantitative agreement with the experimental data shown in Table~\ref{table:dwave}.

Finally, we want to emphasize that, our MQDT approach is supposed to work much better on light alkali atoms while not so good for the heavy ones. The reason for this is two fold: a) light atoms usually have smaller hyperfine splitting energy comparing to heavy ones, which justifies the neglecting of energy dependence in $K^c_{S,T}(\epsilon,\ell)$; (b) the effects of anisotropic interactions including magnetic dipole-dipole interaction and second order spin-orbit coupling are more important in heavy atoms which are not taken into account in our MQDT calculations. As a result, we expect our MQDT to have a similar predicting power on resonances of light atoms like Li, Na, K and their mixtures while it may not be so accurate on heavy ones like Rb and Cs.

Besides the resonance position, several other parameters are required for a complete description of a well isolated resonance. Close to each $\ell$-wave resonance, we have:
\begin{eqnarray}
a_{\ell}(B)=a_{bg\ell}\left(1-\frac{\Delta_B}{B-B_{res}}\right)
\end{eqnarray}
where $a_{\ell}$ is the generalized $\ell$-wave ``scattering length" or the so called scattering hyper-volume and $B_{res}$ is the position of resonance. Another two parameters $\zeta_{res}$ and $\delta\mu$ are defined through the leading order energy and magnetic field dependence of the quantum defect $K_{\ell}^{c0}(\epsilon,B)$ around $\epsilon=0$ and $B=B_{res}$ \cite{Gao2011PRA}:
\begin{eqnarray}
K_{\ell}^{c0}(\epsilon,B)=-w_{\ell}\zeta_{res}^{-1}\frac{\epsilon-\delta\mu\delta B}{E_6}+O(\epsilon^2,\delta B^2,\delta B\epsilon)\label{zetares}
\end{eqnarray}
where $\delta B=B-B_{res}$, $w_{\ell}=[(2\ell+3)(2\ell-1)]^{-1},$ and $E_6=\hbar^2/(m\beta_6^2)=E_{vdW}/4$.

The quantity $\zeta_{res}$ characterize whether the resonance is open-channel ($|\zeta_{res}|\gg1$) or closed-channel ($|\zeta_{res}|\ll1$) dominated. It was shown that $\zeta_{res}>0$ for s-wave and $\zeta_{res}<0$ for all high partial waves \cite{Gao2011PRA}. For a Feshbach resonance, $\delta\mu$ defined in Eq.~\ref{zetares} approximately equals to the magnetic moment difference between the free atom pair and closed channel molecule. However, for a shape resonance, since there is no closed channel molecule, the value of $\delta\mu$ does not have such physical meaning. One can also show that $\zeta_{res},~a_{bg\ell},~\delta\mu$ and $\Delta_B$ are related as:
\begin{eqnarray}
\zeta_{res}=-w_{\ell}\frac{a_{bg\ell}}{\bar{a}_{\ell}}\frac{\delta\mu\Delta_B}{E_6}.
\end{eqnarray}
The above resonance parameters provide a complete description for each isolated resonance and are shown in the Table~\ref{table:swave}-\ref{table:dwave}.

\section{Experimental results}

\subsection{s-wave Feshbach resonances in $^{41}$K-$^6$Li mixture}
We study the $^{41}$K-$^6$Li interspecies s-wave Feshbach resonances for 10 different spin combinations, including 4 elastic channels ($|aa\rangle$, $|ba\rangle$, $|ca\rangle$, $|cb\rangle$) and 6 inelastic ones ($|ab\rangle$, $|ac\rangle$, $|bb\rangle$, $|bc\rangle$, $|cc\rangle$, $|da\rangle$). Fig.\ref{swave} shows a typical inelastic loss spectrum of $^{41}$K for channel $|ca\rangle$, which are fitted with asymmetric Lorentz function $N\propto 1/(4(B-B_0)^2+\omega^2)$, where $\omega=2\omega_0/(1+\exp(F\cdot(B-B_0)))$ with F being the asymmetry parameter. It's quite interesting that the distance between two neighbouring resonances are nearly identical and the resonance width decrease monotonously from low field to high field, which we also find for $|ba\rangle$, $|ca\rangle$, $|ab\rangle$, and $|bb\rangle$ channels, respectively. For all the 10 spin combinations, we successfully find 56 s-wave Feshbach resonances for a magnetic field range between 0~G and 600~G, which are summarized in Table~\ref{table:swave}. The magnetic field accuracy is calibrated by rf spectroscopy between the two lowest hyperfine states of $^{41}$K atoms for several magnetic field values and it is found to be better than 10 mG.

\begin{table}[htbp]
\centering
\begin{tabular}{p{1.4cm}p{1.5cm}p{1.4cm}p{1.8cm}p{1.5cm}}
\hline \hline
Channel&\tabincell{c}{$B_{the}$(G)}&\tabincell{c}{$\Delta_{the}$(G)}& \tabincell{c}{$B_{exp}$(G)}&\tabincell{c}{$\Delta_{exp}$(G)}\\
\hline
$|a,a\rangle$&21.15&0.01&21.17(1)&0.13(1)\\
~           &31.61&0.32&31.63(1)&0.31(2)\\
~           &99.50&0.05&99.56(1)&0.16(1)\\
~           &104.04&0.02&104.10(1)&0.14(1)\\
~           &335.18&0.96&335.13(1)&0.42(1)\\
~           &341.36&0.07&341.29(1)&0.17(2)\\
$|b,a\rangle$&34.34&0.01&34.39(1)&0.14(1)\\
~           &46.29&0.43&46.33(2)&0.23(3)\\
~           &99.30&0.01&99.40(1)&0.14(1)\\
~           &114.84&0.05&114.93(1)&0.15(1)\\
~           &118.96&0.06&119.04(1)&0.16(1)\\
~           &352.46&0.91&352.49(1)&0.72(6)\\
~           &361.52&0.33&361.45(1)&0.45(5)\\
~           &369.45&0.01&369.20(1)&0.14(1)\\
$|c,a\rangle$&35.51&0.01&35.64(6)&0.13(1)\\
~           &62.10&0.04&62.14(1)&0.14(1)\\
~           &124.29&0.55&124.46(1)&0.26(1)\\
~           &128.03&0.05&128.12(1)&0.15(1)\\
~           &374.58&0.49&374.52(1)&0.24(1)\\
~           &384.14&0.12&384.06(1)&0.16(1)\\
~           &393.43&0.42&393.35(1)&0.21(1)\\
$|c,b\rangle$&83.83&0.06&83.90(1)&0.15(1)\\
~           &154.90&0.04&155.00(1)&0.20(1)\\
~           &160.94&0.16&161.04(1)&0.17(1)\\
~           &402.20&0.90&402.23(1)&0.36(23)\\
~           &409.89&0.17&409.83(3)&0.17(2)\\
$|a,b\rangle$&37.57&0.03&37.60(2)&0.69(7)\\
~           &49.27&0.06&49.29(8)&3.81(24)\\
~           &107.06&0.55&107.21(1)&0.26(1)\\
~           &358.90&0.51&358.92(18)&4.68(74)\\
~           &367.56&0.14&367.58(4)&1.21(13)\\
~           &375.58&0.42&375.55(2)&0.30(3)\\
$|a,c\rangle$&78.27&-0.01&78.29(1)&0.26(1)\\
~           &88.09&0.02&88.09(1)&1.24(4)\\
~           &154.18&0.04&154.29(2)&1.93(10)\\
~           &163.95&-0.05&164.04(1)&0.30(3)\\
~           &389.38&0.11&389.21(9)&5.26(38)\\
~           &398.30&0.32&398.27(5)&2.40(16)\\
$|b,b\rangle$&63.86&0.12&63.41(1)&0.21(1)\\
~           &130.25&0.23&130.34(2)&0.74(5)\\
~           &142.70&-0.02&142.77(1)&0.17(1)\\
~           &378.38&0.95&378.37(5)&3.79(15)\\
~           &387.58&0.17&387.52(1)&0.65(3)\\
~           &396.82&0.30&396.70(1)&0.51(4)\\
$|b,c\rangle$&103.98&0.01&104.01(1)&0.52(3)\\
~           &172.08&0.01&172.25(8)&0.54(7)\\
~           &176.53&0.03&176.64(3)&1.06(1)\\
~           &410.54&0.45&410.55(12)&3.92(32)\\
~           &417.74&0.12&417.77(7)&1.58(15)\\
$|c,c\rangle$&196.11&-0.04&196.21(1)&0.14(1)\\
~           &434.16&0.51&434.17(1)&3.13(34)\\
$|d,a\rangle$&102.41&0.82&102.44(1)&0.36(2)\\
~           &166.75&-0.06&166.86(1)&0.43(2)\\
~           &171.33&3.54&171.44(1)&0.48(3)\\
~           &402.36&0.14&402.33(2)&0.95(5)\\
~           &410.02&0.46&409.94(1)&0.61(3)\\
\hline \hline
\end{tabular}
\caption{$^{41}$K-$^6$Li s-wave Feshbach resonance spectroscopy. $B_{the}$ and $\Delta_{the}$ are the theoretical predicted resonance position and width from MQDT calculation. $B_{exp}$ and $\Delta_{exp}$ are the corresponding experimentally determined values. All the above resonances are identified as narrow resonances. The theoretical widths that are less than 0.01 G are recorded as 0.01 G.}
\label{table:swave}
\end{table}

\begin{figure}[htbp]
\includegraphics[width=0.48\textwidth]{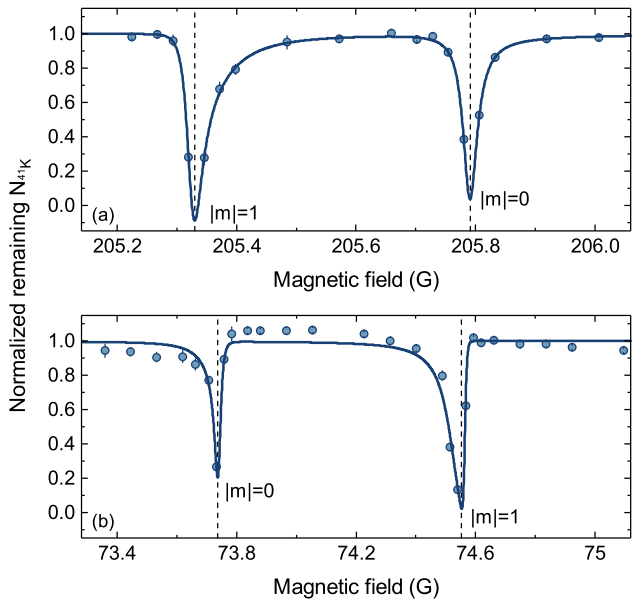}
\caption{Normalized remaining atom number of $^{41}$K as a function of magnetic field, in the vicinity of two p-wave resonances in channel $|cb\rangle$. The solid lines are the multi-peak asymmetric Lorentz fitting curves. The vertical dashed lines mark the fitted resonance positions. We can see obvious doublet structure in both (a) and (b). Because of the different sign of the $\delta\mu$, the distributions of $|m|=0$ and $|m|=1$ resonances are inverted. All the data points are averaged out of five experimental records and the error bars represent the standard deviations.}
\label{pwave}
\end{figure}

\subsection{p-wave Feshbach resonances in $^{41}$K-$^6$Li mixture}

The p-wave resonances $^{41}$K-$^6$Li mixture are measured in the 4 elastic channels $|aa\rangle$, $|ba\rangle$, $|ca\rangle$, and $|cb\rangle$, respectively. Fig.\ref{pwave} shows two typical p-wave Feshbach resonances in $|cb\rangle$ channel. The unambiguous doublet structure of the loss spectrum gives a direct evidence of p-wave resonance, where the two peaks correspond to $|m|=0$ and $|m|=1$ resonances due to the magnetic dipolar interactions~\cite{Ticknor2004PRA}. The large doublet splitting often implies the p-wave resonances are `broad', which are also confirmed by the calculated resonance parameter. Moreover, all of the loss curves display an asymmetric shape. It's because the atoms can access the quasi-bound states that lie above the threshold of the open channel due to their non-zero kinetic energy and thus cause enhanced loss of them. Another intriguing feature is the asymmetric shape inversion of the two loss spectrums. In Fig.\ref{pwave}(a), the ``threshold'' edge of the two resonances appear at the low magnetic field side, while they appear at the high magnetic field side in Fig.\ref{pwave}(b). We attribute this to the different sign of $\delta\mu$. If $\delta\mu>0$ ($\delta\mu<0$), the molecular state moves upward (downward) with respect to the threshold of the open channel with increasing magnetic field, which imply that the quasi-bound states exist at the higher (lower) magnetic field, respectively. Besides, the locations of $|m|=0$ resonances are also inverted due to the different sign of $\delta\mu$, which will be at higher (lower) field for $\delta\mu>0$ ($\delta\mu<0$).

\begin{table*}[htbp]
\centering
\begin{tabular}{p{1.5cm}p{1.5cm}p{1.5cm}p{2cm}p{1.8cm}p{2cm}p{1.8cm}p{1.4cm}p{1.4cm}p{1.4cm}}

\hline \hline
Channel&\tabincell{c}{$B_{the}$\\(G)}&\tabincell{c}{$\Delta_{the}$\\(G)}&\tabincell{c}{$B_{exp}$(G)\\$|m|=0$}&\tabincell{c}{$\Delta_{exp}$(G)\\$|m|=0$}&\tabincell{c}{$B_{exp}$(G)\\$|m|=1$}&\tabincell{c}{$\Delta_{exp}$(G)\\$|m|=1$}&\tabincell{c}{$\zeta_{res}$}&\tabincell{c}{$\delta\mu/\mu_B$}&$a_{bg}/\bar{a}_l$\\
\hline
$|a,a\rangle$&154.23&18.89&154.42(1)&0.17(2)&154.19(1)&0.17(1)&-0.40&1.48&2.64\\
~           &179.64$^*$&0.27&179.64(3)&0.12(5)& & &-0.002&1.89&0.64\\
$|b,a\rangle$&24.66&-18.19&24.26(1)&0.16(1)&24.97(1)&0.17(1)&-0.22&-0.71&3.22\\
~           &150.59&20.25&151.01(1)&0.17(1)&150.43(1)&0.17(1)&-0.30&0.82&3.30\\
~           &187.57$^*$&11.95&187.66(1)&0.13(2)& & &-0.09&1.37&1.04\\
$|c,a\rangle$&25.44$^*$&-53.99&25.60(1)&0.17(1)& & &-0.23&-1.06&0.73\\
~           &65.25&-10.25&64.71(1)&0.15(2)&65.68(1)&0.16(1)&-0.13&-0.72&3.38\\
~           &168.66&11.23&169.19(1)&0.16(2)&168.50(1)&0.16(1)&-0.14&0.65&3.44\\
~           &210.09&-8.52&210.19(1)&0.17(2)&210.06(1)&0.13(3)&-0.14&1.40&2.20\\
~           &226.76&6.73&226.91(2)&0.22(9)&226.77(1)&0.15(1)&-0.06&1.74&0.93\\
$|c,b\rangle$&74.20&-17.10&73.74(7)&0.15(7)&74.55(4)&0.16(2)&-0.30&-1.07&3.00\\
~           &205.40&22.41&205.79(2)&0.15(1)&205.33(1)&0.15(1)&-0.34&0.97&2.92\\
~           &239.71&2.37&239.77(39)&0.14(9)&239.72(20)&0.15(11)&-0.02&1.50&0.86\\
\hline \hline
\end{tabular}
\caption{$^{41}$K-$^6$Li p-wave Feshbach resonance spectroscopy. $B_{the}$ and $\Delta_{the}$ are the theoretical predicted resonance position and width from MQDT calculation. $B_{exp}$ and $\Delta_{exp}$ are the corresponding experimentally determined values. The * mark the resonances where we only observe single loss peak.}
\label{table:pwave}
\end{table*}

The experimental measured 13 p-wave Feshbach resonances are tabulated in Table~\ref{table:pwave}, where 10 'broad' p-wave resonances are identified with doublet structure. For the observed 3 'narrow' p-wave resonances, only one loss peak can be detected, possibly limited by the small splitting and the controlled resolution of our magnetic field.

\begin{figure}[htbp]
\includegraphics[width=0.48\textwidth]{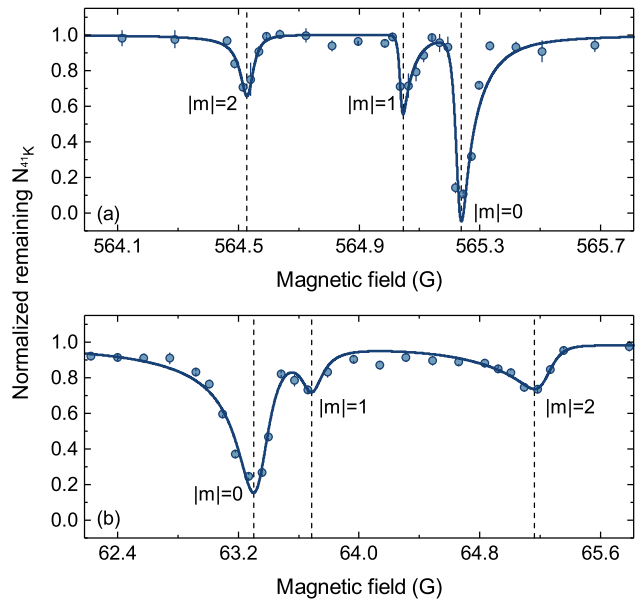}
\caption{Normalized remaining atom number of $^{41}$K as a function of magnetic field, near two d-wave resonances in channel $|bb\rangle$ of $^{41}$K. The solid lines are the multi-peak asymmetric Lorentz fitting curves. The vertical dashed lines mark the fitted resonance positions. We can see obvious triplet structure in both (a) and (b). Because of the different sign of the $\delta\mu$, the distributions of $|m|=0,1,2$ resonances are inverted. All the data points are averaged out of three experimental records and the error bars represent the standard deviations.}
\label{dwave}
\end{figure}

\subsection{d-wave resonance of $^{41}$K atoms}
For this study, we are only interested in the broad d-wave resonances. We successfully find six new broad d-wave resonances in four spin combinations of $^{41}$K atoms ($|aa\rangle$, $|ab\rangle$, $|bb\rangle$, $|cc\rangle$), as shown in Table~\ref{table:dwave}. Together with the one we reported in ref.~\cite{Yao2017}, four of them are shape resonances, while the other three are Feshbach resonances. As an example, Fig.\ref{dwave} shows the loss curves of two d-wave resonances in $|bb\rangle$ channel, where the one at lower magnetic field is a shape resonance. They both display a triplet structure and an asymmetric shape, the hallmark of d-wave resonance.  As the p-wave resonance, three peaks of the d-wave resonance are corresponding to $|m|=0,1,2$ resonances, which is also due to magnetic dipolar interactions. Furthermore, the asymmetric shape inversions are also observed due to the different sign of $\delta\mu$ for this two resonances. Note that for the observed d-wave resonance, the loss of $|m|=0$ resonance is much larger than the $|m|=1,2$ resonances, which is because the dipolar interaction can couple the s-wave scattering state to a d-wave bound state with $|m|=0$, while it's not the case for the p-wave resonance. Moreover, when the $^{41}$K atoms are further cooled to form a pure BEC, i.e., without visible thermal atoms, the two peaks associated with $|m|=1$ and $|m|=2$ resonances disappear. It's because in such temperature, the kinetic energy of atoms is too small to penetrate through the d-wave centrifugal barrier and thus the atomic loss induced by d-wave scattering is negligible. We mention that all the features described above are observed for all the reported d-wave resonances.

\begin{table*}[htbp]
\centering
\begin{tabular}{p{1.3cm}p{1.2cm}p{1.2cm}p{1.6cm}p{1.4cm}p{1.6cm}p{1.4cm}p{1.6cm}p{1.4cm}p{1.15cm}p{1.15cm}p{1.15cm}}

\hline \hline
Channel& \tabincell{c}{$B_{the}$\\(G)}&\tabincell{c}{$\Delta_{the}$\\(G)}&\tabincell{c}{$B_{exp}$(G)\\$|m|=0$}&\tabincell{c}{$\Delta_{exp}$(G)\\$|m|=0$}&\tabincell{c}{$B_{exp}$(G)\\$|m|=1$}&\tabincell{c}{$\Delta_{exp}$(G)\\$|m|=1$}&\tabincell{c}{$B_{exp}$(G)\\$|m|=2$}&\tabincell{c}{$\Delta_{exp}$(G)\\$|m|=2$}&\tabincell{c}{$\zeta_{res}$}&\tabincell{c}{$\delta\mu/\mu_B$}&$a_{bg}/\bar{a}_l$\\
\hline
$|a,a\rangle$&17.79$^{\dag}$&-184.4&16.83(1)&0.44(2)&17.19(6)&0.28(8)&18.75(1)&0.26(5)&-263&-12.22&9.01\\
~           &530.38&4.54&530.48(1)&0.17(3)&530.40(4)&0.13(3)&530.18(2)&0.15(2)&-2.50&1.98&21.47\\
$|a,b\rangle$&25.47$^{\dag}$&1804.3&25.31(2)&0.28(5)&25.41(23)&0.17(40)&25.74(1)&0.43(3)&-258&-10.16&-1.09\\
~           &544.15&3.23&544.93(1)&0.16(1)&544.79(1)&0.16(2)&544.34(11)&0.15(15)&-1.76&1.98&21.25\\
$|b,b\rangle$&64.79$^{\dag}$&-999.4&63.30(1)&0.38(4)&63.69(8)&0.30(16)&65.16(6)&0.45(9)&-16.5&-0.58&2.21\\
~           &564.11&5.56&565.23(4)&0.17(5)&565.05(1)&0.16(5)&564.53(3)&0.15(6)&-0.43&0.30&20.06\\
$|c,c\rangle$&105.66$^{\dag}$&-185.1&104.47(1)&0.43(5)&104.95(1)&0.32(4)&106.27(1)&0.34(3)&-126&-5.99&8.76\\
\hline \hline
\end{tabular}
\caption{$^{41}$K d-wave Feshbach resonance spectroscopy. $B_{the}$ and $\Delta_{the}$ are the theoretical predicted resonance position and width from MQDT calculation. $B_{exp}$ and $\Delta_{exp}$ are the corresponding experimentally determined values. The resonances marked with $\dag$ are broad shape resonances, others are broad Feshbach resonances.}
\label{table:dwave}
\end{table*}

\section{Conclusion and outlook}
In conclusion, we have taken an intensive experimental study on the Feshbach resonances of $^{41}$K-$^6$Li mixture and d-wave Feshbach resonances of $^{41}$K with the aid of MQDT calculation, and confirmed the resonance positions with very high accuracy. Several pretty wide p-wave resonances in $^{41}$K-$^6$Li mixture and extremely broad d-wave resonances in single $^{41}$K system are discovered. Both p- and d- wave resonances have experimental resolvable doublet or triplet structure, respectively. The long background lifetime of the states possessing p- or d- wave resonances enable our system being an ideal platform to pursue many important non-zero partial wave related physics.

\begin{acknowledgements}
We are indebted to valuable discussions with P. Zhang, H. Zhai, and B. Gao. This work has been supported by the National Key R\&D Program of China (under Grant No. 2018YFA0306501, 2018YFA0306502), NSFC of China (under Grant No. 11425417, 11774426), the CAS, and the Fundamental Research Funds for the Central Universities (under Grant No. WK2340000081). R. Qi is supported by the Research Funds of Renmin University of China (under Grants No. 15XNLF18, 16XNLQ03).
\end{acknowledgements}


\begin{thebibliography}{36}
\bibitem [{res()}]{respos}%
  \BibitemOpen
  \href@noop {} {}\bibinfo {note} {According to the Wigner-Eckart theorem, the
  resonance positions without dipole-dipole interaction stastify
  $({B}_0+2{B}_1)/3$ (p-wave) and $({B}_0+{B}_2)/2$ (d-wave), respectively,
  where ${B}_{0,1,2}$ is the resonance position of $|m|=0,1,2$.}\BibitemShut
  {Stop}%
\end{thebibliography}
%

\end{document}